\documentclass{emulateapj}
\newcommand{\til}{$\sim$}
\newcommand{\ergsqcmsec}{\thinspace\hbox{$\hbox{erg}\thinspace\hbox{cm}^{-2}
                \thinspace\hbox{s}^{-1}$}}
\newcommand{\ergsqcmsecA}{\thinspace\hbox{$\hbox{erg}\thinspace\hbox{cm}^{-2}
                \thinspace\hbox{s}^{-1}\thinspace\hbox{\AA}^{-1}$}}
\newcommand{\ergsec}{\thinspace\hbox{$\hbox{erg}\thinspace\hbox{s}^{-1}$}}
\def\spose#1{\hbox to 0pt{#1\hss}}
\def\simlt{\mathrel{\spose{\lower 3pt\hbox{$\mathchar"218$}}
     \raise 2.0pt\hbox{$\mathchar"13C$}}}
\def\simgt{\mathrel{\spose{\lower 3pt\hbox{$\mathchar"218$}}
     \raise 2.0pt\hbox{$\mathchar"13E$}}}

\newcommand{\msun}{\thinspace\hbox{$M_{\odot}$}}
\newcommand{\ta}{SDSS~J0932}
\newcommand{\tb}{SDSS~J1023}
\newcommand{\sat}{{\em XMM-Newton}}

\def\today{\ifcase\month\or
January\or February\or March\or April\or May\or June\or
July\or August\or September\or October\or November\or December\fi
\space\number\day, \number\year}

\slugcomment{Accepted for publication in the Astronomical Journal 2005 September 23}

\shorttitle{SDSS J0932 and SDSS J1023}
\shortauthors{Homer et al.}

\begin{document}


\title{{\em XMM-Newton} and optical follow-up observations of SDSS~J093249.57+472523.0 and SDSS~J102347.67+003841.2$^{\dagger,\star}$}


\altaffiltext{$\dagger$}{Some of the results presented here were obtained with the MMT
Observatory, a facility operated jointly by The University of Arizona and the
Smithsonian Institution.}
\altaffiltext{$\star$}{Based on
observations obtained with the Sloan Digital Sky Survey and with the
 Apache Point
Observatory (APO) 3.5m telescope, which are owned and operated by the
Astrophysical Research Consortium (ARC)}
\author{Lee Homer\altaffilmark{1}, Paula Szkody\altaffilmark{1}, Bing Chen\altaffilmark{2,3}, Arne Henden\altaffilmark{4,5,6}, Gary Schmidt\altaffilmark{7}, Scott
  F. Anderson\altaffilmark{1}, Nicole M. Silvestri\altaffilmark{1} and J. Brinkmann\altaffilmark{8}}
%\affil{Apache Point Observatory, 2001 Apache Point Road, P.O. Box 59, Sunspot, NM 88349-0059}
%\email{jb@apo.nmsu.edu}

\altaffiltext{1}{Department of Astronomy, University of Washington, Box 351580, Seattle, WA 98195, USA}
\email{homer@astro.washington.edu}
\altaffiltext{2}{\sat\ Science Operations Centre, ESA/Vilspa, 28080, Madrid, Spain}
\altaffiltext{3}{VEGA IT GmbH, c/o European Space Operations Centre, Darmstadt, Germany}
\altaffiltext{4}{Universities Space Research Association}
\altaffiltext{5}{US Naval Observatory, Flagstaff Station, P.O. Box 1149, Flagstaff, AZ 86002-1149, USA}
\altaffiltext{6}{American Association of Variable Star Observers, 25 Birch Street, Cambridge, MA 02138, USA}
\altaffiltext{7}{The University of Arizona, Steward Observatory, Tucson, AZ 85721, USA}
\altaffiltext{8}{Apache Point Observatory, 2001 Apache Point Road, P.O. Box 59, Sunspot, NM 88349-0059, USA}
%\topmargin -0.5in
%\textheight 9.20in

%\received{2004}
%\accepted{2004}

%==============================================================================

\begin{abstract} We report follow-up \sat\ and ground-based optical 
observations of the unusual X-ray binary SDSS~J102347.67+003841.2 
(=FIRST~J102347.6+003841), and a new candidate intermediate polar found in 
the Sloan Digital Sky Survey: SDSS~J093249.57+472523.0. \tb\ was observed 
in its low-state, with similar magnitude/color ($V=17.4$ and $B=17.9$), 
and smooth orbital modulation as seen in most previous observations. We 
further refine the ephemeris (for photometric minimum) to: ${\rm 
HJD(TT)_{min}}= 2453081.8546(3) + E* 0.198094(1){\rm d}$. It is easily 
detected in X-rays at an unabsorbed flux (0.01-10.0 keV) of 
$5\times10^{-13}$\ergsqcmsec.  Fitting a variety of models we find that: 
(i) either a hot ($kT\simgt15$ keV)  optically thin plasma emission model 
(bremsstrahlung or {\sc MEKAL}) or a simple power law can provide adequate 
fits to the data; (ii) these models prefer a low column density 
\til$10^{19}$ cm$^{-2}$; (iii) a neutron star atmosphere plus power law 
model (as found for quiescent low-mass X-ray binaries) can also produce a 
good fit (for plausible distances), though only for a much higher column 
$\approx4\times10^{20}$ cm$^{-2}$ and a very cool atmosphere ($kT\simlt50$ 
eV).  These results support the case that \tb\ is a transient LMXB, and 
indeed places it in the subclass of such systems whose quiescent X-ray 
emission is dominated by a hard power law component.  Our optical 
photometry of \ta\ reveals that it is an high inclination eclipsing 
system.  From our two epochs of data, and 7 eclipse times, we are able to 
derive a best fit ephemeris for minimum light: ${\rm HJD(TT)_{min}}= 
2453122.2324(1) + E* 0.0661618(4) {\rm d}$, although aliases, with one 
cycle count different between epochs, are acceptable.  The X-ray spectrum 
is well fit by either a hard bremsstrahlung or power law, with a partial 
covering absorption model, with a high covering fraction \til0.9 and 
column $\approx10^{23}$ cm$^{-2}$.  Combined with its optical 
characteristics --- high excitation emission lines, and brightness, 
yielding a large $F_X/F_{opt}$ ratio --- this highly absorbed X-ray 
spectrum argues that \ta\ is a strong IP candidate.  However, only more 
extensive optical photometry and a detection of its spin or spin-orbit 
beat frequency can confirm this classification. %\vspace*{4cm} 
\end{abstract} 

\section{Introduction}
Cataclysmic variables (CVs) and low-mass X-ray binaries (LMXBs) are two broad classes of interacting binary star, whose emission, from X-ray
through optical/IR (and even radio), is powered by accretion onto a compact object (white dwarf or neutron star/black hole respectively).  There are good reviews of the various sub-classes and general characteristics in \citet{warn95}, and
\citet{lewi95} respectively. When actively accreting, both classes exhibit  (broad) emission line spectra, with
very blue continua (from the accretion disk), but are normally distinguished by their relative X-ray output; their
X-ray luminosity roughly scales as $M_{\ast}/R^{2}_{\ast}$, i.e. a factor of $10^{4}-10^5$. However, in many cases the accretion rate onto the compact
object is highly variable, with both CVs and LMXBs exhibiting a variety of brightness states.

For the past three years, we have used \sat\ to study and confirm potential magnetic CVs identified in the Sloan
 Digital Sky Survey \citep[SDSS]{york00}. Two SDSS sources stood out based on the unusual strength of their He emission lines, as well as on the double-peaked
 character of all their lines. These are  SDSS~J102347.67+003841.2 \citep{szko03} and SDSS~J093249.57+472523.0 \citep{szko04a}.

SDSS~J102347.67+003841.2 = FIRST J102347.6 +003841(hereafter \tb) was identified as a CV,
 from its radio emission in the
FIRST radio survey \citep[][, the first CV to be so discovered]{bond02} , and independently from  its color and emission line spectrum in the SDSS. In both cases, a highly magnetic white dwarf was suggested to account for the high-excitation
 optical emission lines and detectable radio emission.  The nature of the lines appeared more consistent with an intermediate
 polar (IP), since they lacked both the large equivalent widths and narrow components usually seen in polars.  Significant changes in state have been postulated to account for: (i) very different optical photometric behaviour seen by \citet{bond02} and
 later \citet{woud04}, a change from pure flickering behaviour to a smooth periodic modulation on a period of 4.75 hr; (ii) significant color
 changes between the SDSS photometry (when it appeared red) to the later SDSS spectroscopy, showing a very blue continuum. 

 Most recently,
 \citet{thor05} (hereafter TA05) presented the results of a photometric and spectroscopic campaign, which have called into question whether \tb\ is a CV,
 or a LMXB instead. The low-state spectrum reveals a mid-G star, unlike either previous spectral observation.  Multi-color photometry showed a
 smooth orbital modulation similar to \citet{woud04}, with color changes consistent with the changing aspect of the heated face of the secondary, but even at
 maximum light no emission lines are present in the spectrum. Modelling of the light curves, together with the radial velocity constraints,
 indicates that a high primary mass (greater than the Chandrasekhar mass) is needed, thus making \tb\ a LMXB. We obtained X-ray, circular
 polarization and other optical data to further elucidate the nature of this source. 

For \ta\ we also obtained contemporaneous optical light curves, spectra and X-ray/optical data from \sat\, in an attempt to discern its nature
and possible relation to \tb.

\section{Observations}

\begin{deluxetable*}{lllcll}
%\tablenum{1}
\tablewidth{0pt}
\tabletypesize\small
\tablecaption{Observation Summary\label{tab:obslog}
}
\tablehead{\colhead{SDSS J}&
\colhead{UT Date} & \colhead{Obs} & \colhead{UT Time} & \colhead{Characteristics\tablenotemark{a}} &
\colhead{Comments} }
\startdata
1023&2002 May 11 &  Bok:SPOL & 03:54 - 04:17 & circ. pol.$<0.5\%$ &  spectropolarimetry, 1200s\\
&2004 Feb 16 &  MMT:SPOL & 08:40 - 09:02 & circ. pol.$<0.03\%$ & spectropolarimetry, 1200s\\
&2004 May 12& \sat: &&&\\
&& EPIC-pn & 10:44 -- 14:37 &0.08 cts s$^{-1}$&12570s live time\tablenotemark{b} \\
 && EPIC-MOS1/2 & 10:21 -- 14:42 & 0.03 cts s$^{-1}$&15430s live time \\
 && OM & 10:30 -- 14:14 & $B=17.9$ &12899s duration\\
&2004 May 13 & NOFS & 03:12 -- 06:55 &$\sim V=17.3-17.6$ & open filter photometry\\
&2004 May 23 & APO: DIS &  05:40 -- 06:05   &--- &spectrum (poor fluxing)\\
 
0932 &2004 Apr 11 & NOFS & 03:53 -- 08:48 & $\sim V=18.5-19.0$&open filter photometry\\
     &2004 May 12 & NOFS & 03:08 -- 06:41 & $\sim V=18.5-19.2$&open filter photometry\\
&2004 May 12 & \sat:  &&&\\
&&EPIC-pn & 07:08 -- 08:57 & 0.008 cts s$^{-1}$&5698s live time \\
& &  EPIC-MOS1/2 &  06:46 -- 02:00 & 0.003 cts s$^{-1}$&7953/7984s live time \\
 &&  OM & 06:54 -- 09:04 & $B=19.1$ &7478s duration\\
&2004 May 23 & APO: DIS &  05:08 -- 05:33 & $\sim V=19.3$ &spectrum \\

\enddata
\tablenotetext{a}{The open filter photometry from NOFS has an estimated
  equivalent $V$ zero-point, while for spectra the flux density at \til5500\AA\ is used. The \sat\ count rates are average values for each
  observation for a single detector.}
\tablenotetext{b}{The live time of the X-ray CCD detectors refers to the sum of the good-time intervals, less any dead time.  It
  is typically much less than the difference of observation start and stop times.}
%\tablenotetext{c}{The absolute fluxing of the spectra for \tb\ are unreliable since it was observed at very high airmass, hence we do not quote an equivalent $V$ magnitude.}

\end{deluxetable*}
A summary of the X-ray and optical observations is presented in Table~\ref{tab:obslog}.
\sat\ possesses three X-ray telescopes, backed by the two MOS and one pn \citep{turn01} CCD cameras, and an optical monitor \citep[OM,
][]{stru01}; hence parallel data are obtained by all detectors.  The two Reflection Grating Spectrographs \citep{denH01} are arrayed in the optical path of the MOS detectors; however, for neither target were there sufficient counts to
 provide useful signal in the resulting dispersed
spectra.  In the direct (spectro)imaging EPIC detectors count rates were sufficient for
extraction of low-resolution spectra and light curves. We followed the standard protocol as given by the Vilspa \sat\ Science
Analysis System (SAS) site\setcounter{footnote}{8}
\footnote{Available from
  http://xmm.vilspa.esa.es/external/ xmm\_sw\_cal/sas.shtml} and the ABC
guide\footnote{http://heasarc.gsfc.nasa.gov/docs/xmm/abc/abc.html} from the US GOF. In both cases we used calibration files current to 2005
March 16 and the SAS v6.1.0, and as precaution we used SAS-tools to produce new event list files from the Observation Data Files incorporating the latest calibration updates.

  To check on variations in the X-ray background we created light curves for each entire detector in the 10--15~keV range. No hard X-ray flaring events
  were present in the data, and for \tb\ we simply used the standard good time intervals (GTIs).  The case of \ta\ was more complex. Due to its low count rate, the
  effects of background are relatively important. Examination of the background
  spectra showed the usual steep rise at low energies (due to detector noise) and a number of prominent metal lines, which are due to
  fluorescence of satellite materials bombarded by soft protons, especially in the pn.  For the pn it is possible to make some improvements to
  reduce this detector
  background; we followed the guidelines in  SAS  handbook section 4.3.2.1 ``Improving the quality of EPIC pn data: epreject.''  Next, 
  we created background light curves to examine any variations at these specific energies, and identified one interval of especially high flux, and created new
  GTIs to exclude it.

In addition to the time selection,  we  screened the event lists using the standard canned expressions
and restricted energies to the 0.1--10~keV range.   Owing to the very different count rates, we used different extraction radii for each
  source.  For \ta\ we used  circular source apertures of radius 320~pixel (MOS) and 360~pixel (pn), in each case enclosing $\sim70\%$ of
the energy, chosen to maximize S/N.  For \tb\ we increased the radii to 480 and 640~pix respectively, now enclosing $\sim80\%$ of
the energy.  For the background, we selected annuli on the same chip for the MOS, and  adjacent rectangular regions at similar detector Y
  locations to the target for the pn, excluding any point sources.  As advised we selected single to quadruple events for the MOS (pattern
  $\leq$ 12), but only singles from the pn (pattern = 0).

Using the same extraction regions for source and background we extracted their light curves with SAS task {\tt evselect}.  In this instance (to
maximise counts) we included both singles and doubles from the pn (pattern$\leq4$). Finally, background subtraction (scaled by extraction area)
and correction of the time stamps to the heliocenter (times in HJD(TT)\footnote{The tools actually yield barycentric Julian Date in the barycentric dynamical time system, BJD(TB).  However, the
  offset to heliocentric Julian Date in the geocentric (terrestrial) dynamical time system (HJD(TT)) is less than \til3~s at any given time, fine for our
  purposes here.}) were undertaken with FTOOLS\footnote{http://heasarc.gsfc.nasa.gov/lheasoft/ftools/}. We obtained simultaneous $B$-band light curves from the OM.  Again, we performed our own extractions using {\tt omfchain} starting from the ODF event list files,
setting our binning times at this earliest stage. We converted the count rates into magnitudes according to the results from interactive
photometry with the {\tt omsource} task.

\ta\ was also observed by the
US Naval Observatory Flagstaff Station (NOFS) 1m telescope on 2005 April 11 and May 12.  The latter observation covers the 3 hr window immediately
preceding the \sat\ observation (and OM optical coverage).  To maximise signal-to-noise these
data were taken in white light, but an approximate zeropoint for this differential photometry to Johnson $\sim V$-band  was made possible through calibration of the field from all-sky photometry
including Landolt standards observed at NOFS. Hence, the light curves are labelled $~V$ magnitudes. A night after its X-ray observations (2005 May 13) a similar time
series was taken on \tb. Circular spectropolarimetry and spectrophotometry was also obtained for \tb\ with the CCD Spectropolarimeter SPOL
(Schmidt et al. 1992) in 2002 May
on the 2.3~m Kitt Peak Bok telescope and on the 6.5~m MMT in 2004 February, as
indicated in Table 1.  At both epochs, the object displayed the spectrum of
a G star, with apparent magnitudes of 17.9 and 17.5, respectively.  The 20~min
observation sequences revealed no significantly polarized features, and
respective upper limits are $v=V/I=0.5$\% and 0.03\% when summed over the
spectrum.

\begin{figure*}[!tb]
%\resizebox{.95\textwidth}{!}{\rotatebox{0}{\plottwo{0932_APOspec.eps}{1023_APOspec.eps}}}
\resizebox{.95\textwidth}{!}{\rotatebox{0}{\plottwo{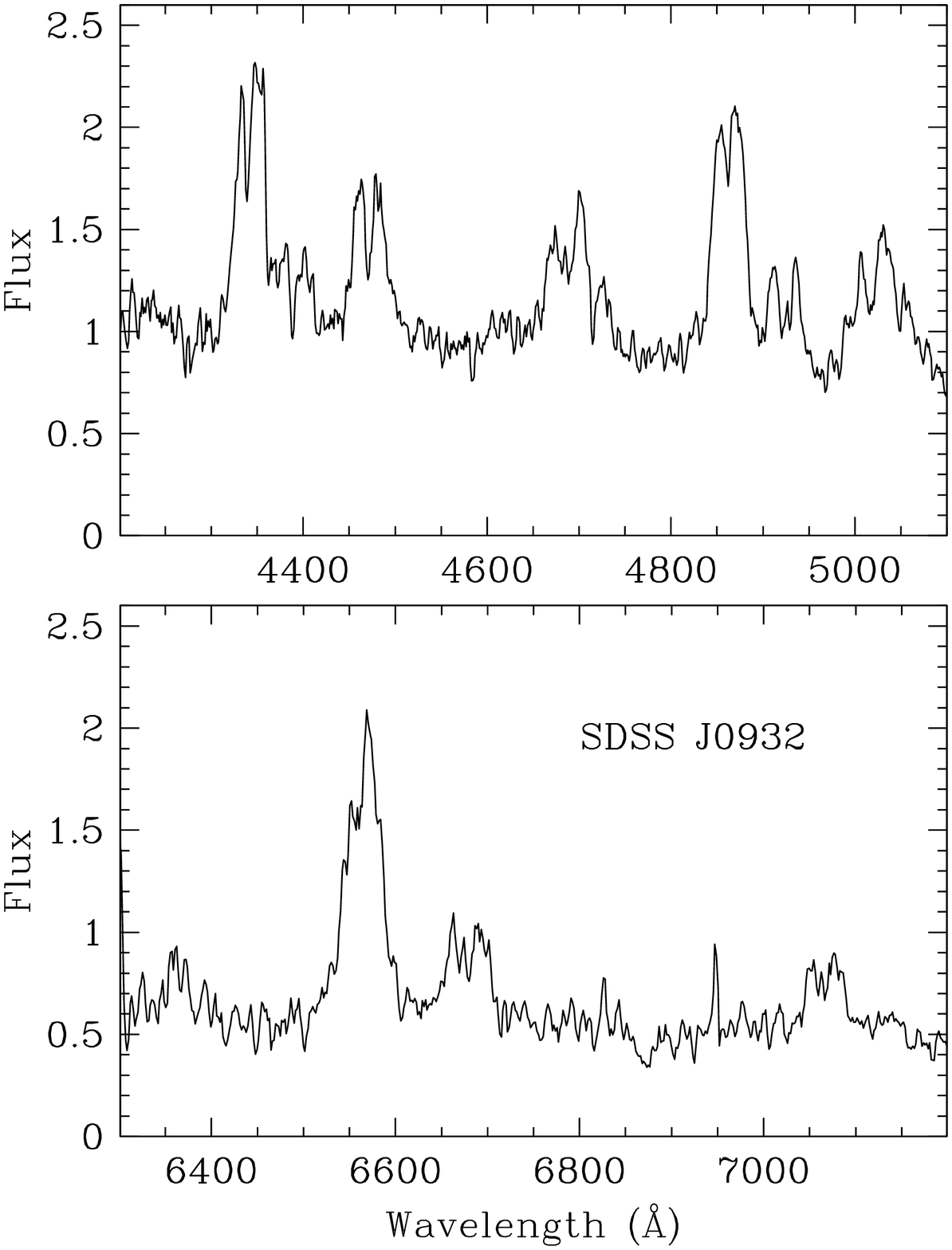}{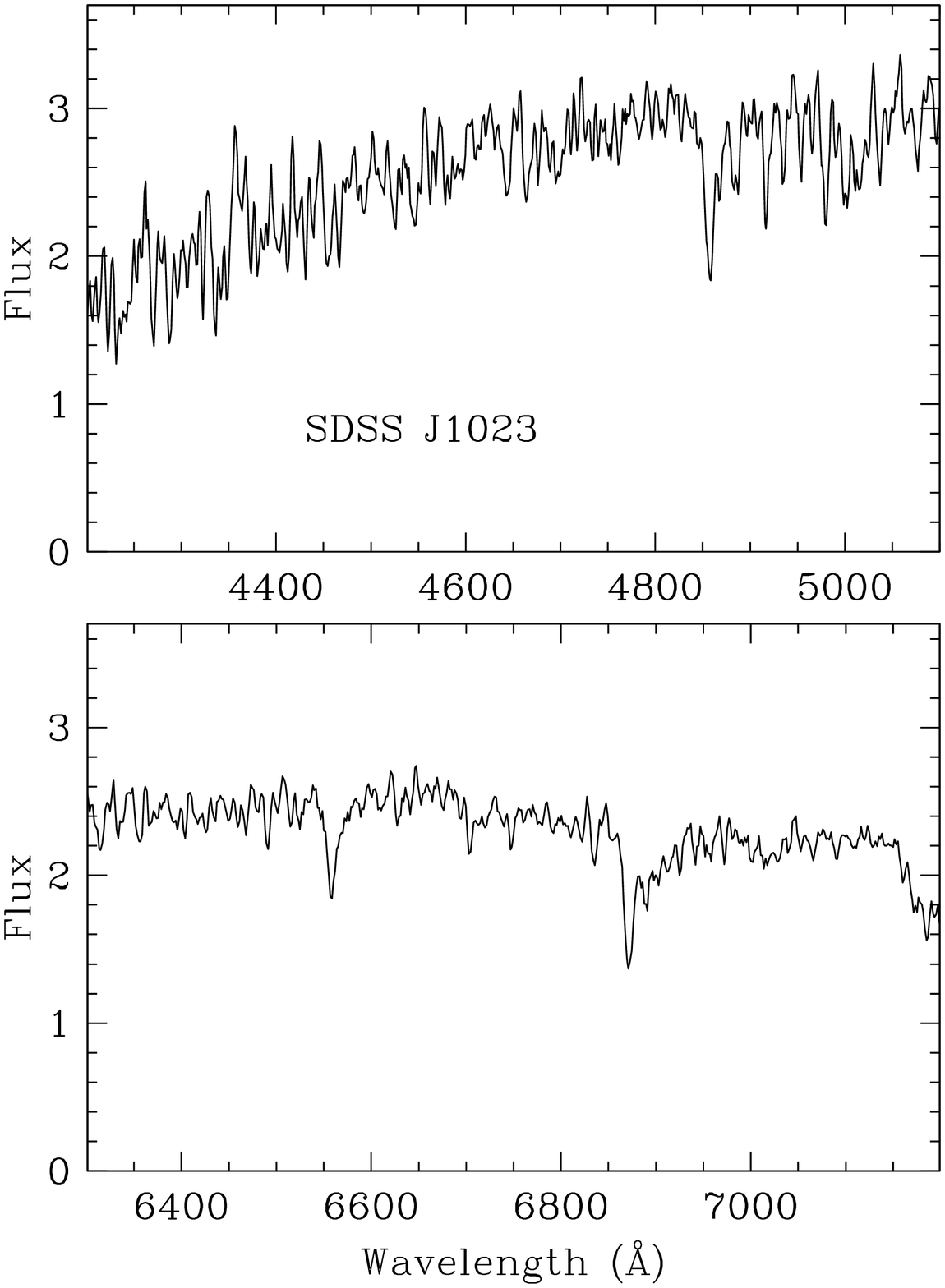}}}
\caption{APO spectra of \ta\ ({\it left panels}) and \tb\ ({\it right panels}).  \ta\ was observed in a similar state to that seen during its SDSS spectroscopy; it exhibits strong He lines
  indicative of high-ionization states seen in magnetic CVs. In contrast, \tb\ was observed in its low state, showing neither blue light from
  an accretion disc nor any emission lines. The
  flux scale is in units of flux density $10^{-16}$\ergsqcmsecA. \label{fig:optspec}}
\end{figure*}

We also observed both \ta\ and \tb\ on 2005 May 23 with the double-imaging spectrograph (DIS) with a resolution of about 2\AA\ on the Apache Point
Observatory (APO) 3.5m (see fig.~\ref{fig:optspec}).  These spectra were reduced and flux-calibrated using standard IRAF\footnote{IRAF (Image Reduction and Analysis Facility) is distributed by the National Optical Astronomy Observatories, which are operated
  by the Association of Universities for Research in Astronomy (AURA) Inc., under cooperative agreement with the National Science Foundation} routines.

\section{\tb}

\subsection{Light Curves}
\begin{figure}[!tb]
%\resizebox{.45\textwidth}{!}{\rotatebox{0}{\plotone{1023_fds_col.eps}}}
\resizebox{.45\textwidth}{!}{\rotatebox{0}{\plotone{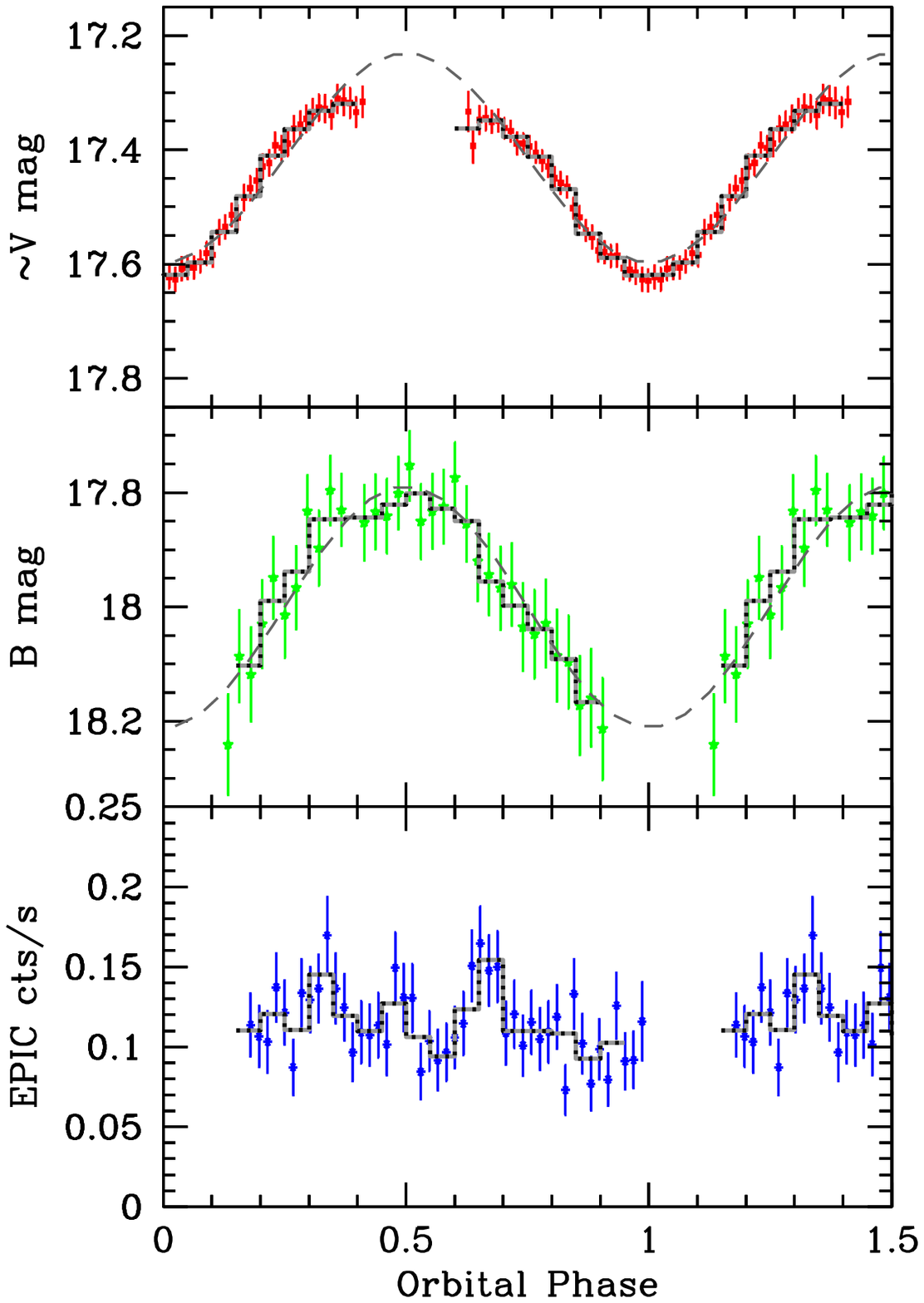}}}
\caption{\tb: from top to bottom, light curves in white light from NOFS, $B$-band from the \sat/OM and in the X-ray (0.2-10 keV).  They are plotted
  against phase according to our refined ephemeris. The optical curves show smooth, but non-sinusoidal profiles, as can be seen by the
  deviations from the overplotted best fit sinusoids.\label{fig:1023fds}}
\end{figure}

Figure~\ref{fig:1023fds} shows that when we combine our two optical
light curves (phase-folded according to our refined
ephemeris), we have
full phase coverage.  Again we see the same kind of smooth (but
non-sinusoidal) modulation as observed by both \citet{woud04} and TA05.  Both our optical curves indicate a broad (\til0.3
in phase) constant flux interval at maximum light, followed by a smooth rounded minimum.  Excellent agreement is seen for the phasing of minimum light
(posited as superior conjunction of the donor); 574 cycles have elapsed between the TA05 ephemeris and our epoch of observation,
giving a phase uncertainty of merely 0.006.  Fitting a Gaussian to the minimum in the 2004 May 13 NOFS lightcurve we determine
$T_0'=2453138.7077(3)$, giving $P'=\Delta(T_0)/574=113.7062(6)/574=0.198094(1)$d, and a refined ephemeris of:
\[ {\rm HJD(TT)_{min}}= 2453081.8546(3) + E* 0.198094(1) {\rm d}\] 

Note our ephemeris is given in Terrestrial Time  (= UT + 64.184s at the current epoch\footnotetext{We use Terrestrial Time throughout this
  paper, since it the system best suited for ephemerides, unlike UT which is adjusted by leap seconds as needed to match the earth's rotation}).

We constructed the X-ray light curve (Fig.~\ref{fig:1023fds}) by summing the background-subtracted light curves from all three EPIC cameras, with 300s bins.  The X-ray light curve does not show a similar smooth modulation, but instead is quite irregular.  Indeed, the amplitude of its variability
(peak-peak) amounts to about 60\%, greater than the 40\% in $B$ or 30\% in white
light. In view of the IP nature hypothesis, we note that it is possible to fit a sinusoid plus first harmonic model to our data, with a period of $0.069\pm0.003$ d
(98 min), or $0.3\times P_{\rm orb}$. As to the significance of this signal, the peak in the
Lomb-Scargle periodogram \citep{scar82} is at the 96.5\% confidence level, but we caution  that with merely two cycles of coverage its reality
remains debatable.  Moreover, only three IPs have similarly large spin-orbit period ratios, V1025 Cen, DW Cnc and EX Hya, and all these are IPs
below the period gap, with periods \til1.5 hrs.  If this putative spin period were to be confirmed, it would place \tb\ in a unique position in
the $P_{\rm orb}-P_{\rm spin}$ diagram.
%: there are no other confirmed IPs with a similar orbital period $\simgt3$ 3 hr that also have  $P_{\rm spin}/P_{\rm orb}\simgt0.1$.

%the fit semi-amplitudes are $0.015\pm0.004$ cps and  $0.013\pm0.004$, while the peak in the

\begin{deluxetable*}{llllll}
%\tablenum{1}
\tablewidth{0pt}
\tablecaption{X-ray Spectral Fits for \tb\label{tab:1023xfits}}
\tablehead{ \colhead{Model} &\colhead{reduced } & \colhead{$N_H$} &\colhead{$kT$\tablenotemark{a}} &\colhead{$kT$ or $\Gamma$} & \colhead{Flux\tablenotemark{b}}\\
  & \colhead{$\chi^2$} & \colhead{$\times10^{20}$cm$^{-2}$}&(eV)& (keV) &}

\startdata
 bremss & 1.45 & 4.22\tablenotemark{c} &\nodata &$15^{+4}_{-2}$ &$4.9\pm0.4$ \\
             & 1.06 &  0.1 & \nodata&$63^{+26}_{-30}$ & $5.2\pm0.8$\\
    {\sc mekal} & 1.38 &  4.22 &\nodata &$12^{+3}_{-2}$ & $5.1\pm0.5$\\
           & 1.02 &  0.1 &\nodata &$39^{+24}_{-12}$ & $5.1\pm0.6$\\
    PL  & 1.16 & 4.22 &\nodata &  $1.47\pm{0.04}$ & $5.5\pm0.3$ \\
        &  1.02 & 0.1 &\nodata &  $1.27\pm{0.03}$ &  $5.3\pm0.3$\\
	     & 1.02 & $0.9\pm0.7$ &\nodata & $1.31\pm0.05$ &  $5.3\pm0.3$\\ 
   2 T {\sc mekal} & 1.09 &  4.22 & $98^{+6}_{-4}$ &$23^{+14}_{-7}$ &$11\pm4$ \\
        & 0.99 &  0.1 & $300^{+24}_{-12}$ & $80^{+{\rm pegged}}_{-35}$ &$5.2^{+0.2}_{-0.4}$ \\
   BB + {\sc mekal}&  1.00 &  4.22 & $130\pm{20}$ &$80^{+{\rm pegged}}_{-42}$ &  $5.5\pm0.3$\\
        & 0.98 &  0.1 & $220\pm{40}$ & $80^{+{\rm pegged}}_{-19}$ & $5.0\pm0.3$ \\
PL + NSA:\hfill 1 kpc &1.02 & 4.22              & $33\pm{1}$ &$1.32\pm0.05$         &$6.0\pm0.6$\\
                &1.02 &$0.8^{+3.2}_{-0.7}$& $<21$      &$1.32^{+0.04}_{-0.05}$&$5.2\pm0.5$\\
\hfill          2 kpc &1.00 & 4.22              & $45\pm{2}$ &$1.27^{+0.05}_{-0.06}$&$5.8\pm0.6$\\
                &1.01 &$4.3^{+1.8}_{-2.1}$& $45\pm{5}$ &$1.24^{+0.07}_{-0.04}$&$5.9\pm0.6$\\
\hspace*{1.8cm}         2.3 kpc &0.99 & 4.22              & $48\pm{2}$ &$1.26^{+0.05}_{-0.06}$&$5.8\pm0.6$\\
                &1.02 &$4.4^{+1.6}_{-1.4}$& $48\pm{6}$ &$1.32^{+0.06}_{-0.04}$&$5.8\pm0.6$\\

\enddata
\tablenotetext{a}{For neutron star atmosphere (NSA) model the effective temperature (for a distant observer) is quoted, i.e. $kT_{\rm eff}=g_r\times T_{\rm eff}$ where $g_r=(1-2.952\times M_{\rm ns}/R_{\rm ns})^{0.5}$ is the gravitational redshift parameter.}
\tablenotetext{b}{Unabsorbed flux in the 0.01--10 keV range, in units of $10^{-13}$\ergsqcmsec.}
\tablenotetext{c}{Model parameters without quoted uncertainties have been fixed at the value given.}
\end{deluxetable*}

\subsection{X-ray Spectral Fitting}
Following the most recent guidelines on the low-energy EPIC calibrations, we restricted our fitting to $>0.2$keV for the MOS and $>0.15$keV for the pn, whilst still taking care to
check the effects of different choices of these limits. After initial inspection of the data, showing a high energy tail, we used all good data
below 15 keV for this relatively bright source. There were sufficient counts to bin the data at $>$20 counts/bin, and use $\chi^2$ statistics to
find the best fits to the background subtracted source spectra.  We also double-checked our results using $>$5 counts/bin, and fitting simultaneous source and background spectra using
Cash statistics; in all cases the results were consistent with the  $\chi^2$ approach.

\begin{figure}[!tb]
%\resizebox{.47\textwidth}{!}{\rotatebox{0}{\plotone{1023_xspec_fit.eps}}}
\resizebox{.47\textwidth}{!}{\rotatebox{0}{\plotone{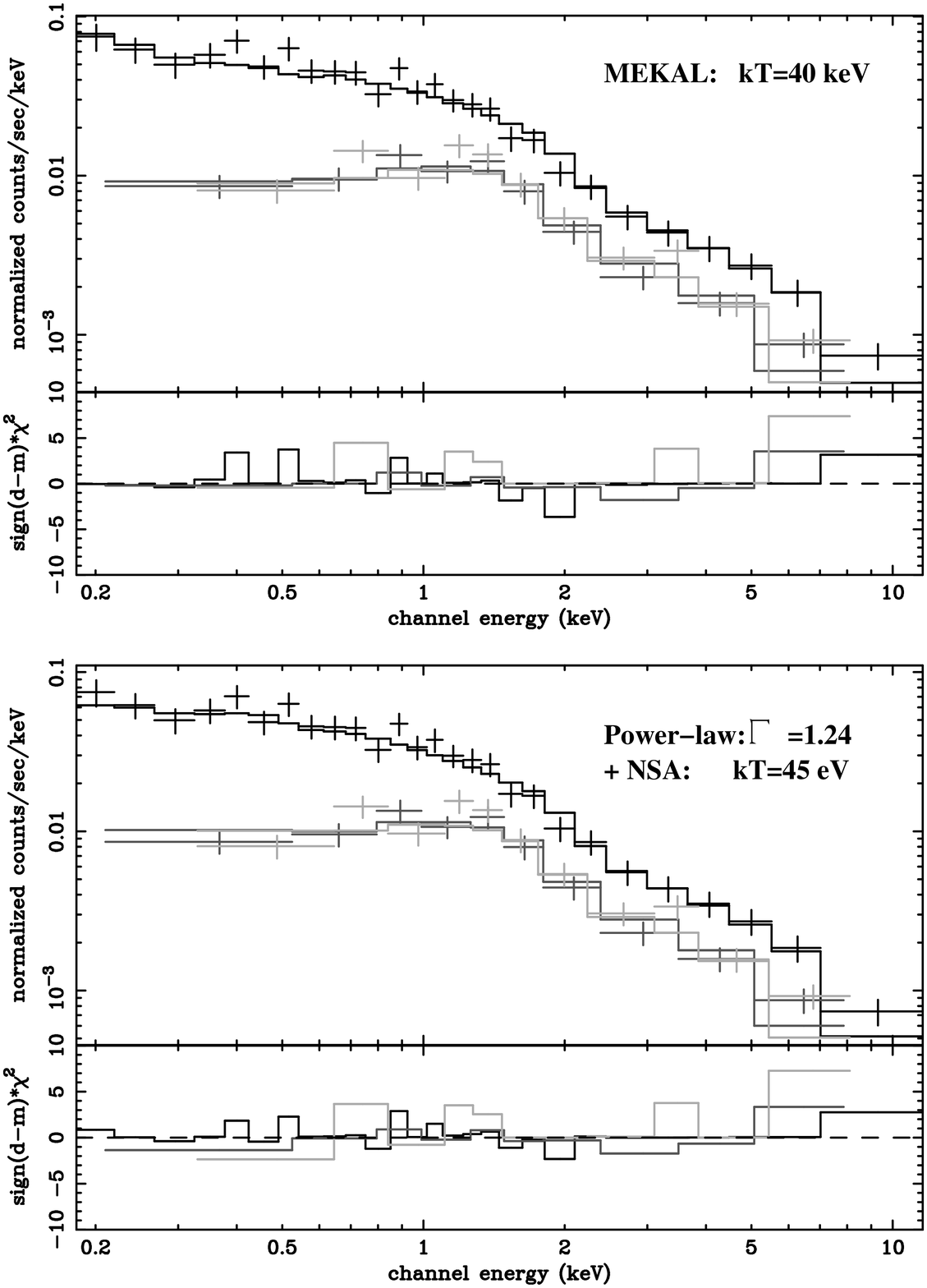}}}
\caption{\sat/EPIC spectra for \tb: {\it black}--pn; {\it dark grey}--MOS1; {\it light grey}--MOS2.  {\it Top}: best fit ``CV'' model
  consisting of a single temperature optically thin plasma model ({\sc mekal}), absorbed by $N_H=1\times10^{19}$ cm$^{-2}$. {\it Bottom}: best
  fit ``quiescent LMXB'' model comprising  a soft neutron star atmosphere (NSA) and a simple power law, with a higher column of $N_H=4.3\times10^{20}$
  cm$^{-2}$. \label{fig:1023xspec}}
\end{figure}

A variety of models were fit to the data, starting with simple single component plus interstellar absorbing column (see Table.~\ref{tab:1023xfits}). In general the emission of
CVs can be approximated with a thermal plasma. We performed fits setting
$N_H=4.22\times10^{20}$ cm$^{-2}$, the maximum Galactic column as derived from dust maps (using HEAsoft tool {\tt nH}) or a fiducial
$N_H=1\times10^{19}$ cm$^{-2}$, appropriate for such a high Galactic latitude source, and lastly with this parameter free. A high temperature ($\simgt30$keV) thermal plasma (either
bremsstrahlung, or bremsstrahlung plus lines-- {\sc mekal}) can produce an acceptable fit (see fig.~\ref{fig:1023xspec}) but only with the lower column option (leaving the column free
yielded only an upper limit of $N_H=4.4\times10^{19}$ cm$^{-2}$, but was consistent with zero). The best fit was a simple power law with
$\chi^2_{\nu}\approx1.0$ for either fixed column, and photon index of 1.3--1.5 (also see fig.~\ref{fig:1023xspec}).  In this case, the data could roughly constrain the column, when freed, to
$N_H=9\pm7\times10^{19}$ cm$^{-2}$.   A power law fit is suggestive of that found for the hard component of LMXBs in quiescence \citep[see e.g.][]{camp98}.

Moving onto more complex models, we examined two temperature and multi-temperature {\sc mekal}
(XSPEC:{\sc cemekl}) models--used to fit IP spectra, and lastly a soft blackbody + {\sc mekal} as for a polar. A two temperature fit was possible,
but in the case of the low column the higher temperature {\sc mekal} pegged at its upper limit of 80 keV, whilst the cool plasma settled at 100--300
eV. The data were not suitable to
constrain the {\sc cemekl} model at all. Adding in a soft blackbody to a {\sc mekal} fit has a similar effect as the two temperature model.   We found
well-constrained $kT_{\rm BB}=130$ and 220 eV respectively for the high and low column cases, but in both cases $kT_{\rm MEKAL}=80$keV (pegged).  This
compares to a typical polar fit with \til30eV and 30 keV for these components.  

On the other hand if \tb\ is a LMXB, then we might expect a good
fit for the typical two component neutron star
atmosphere (NSA) plus power law model.  Fixing both the neutron star radius and mass at their canonical 10 km and 1.4\msun\ values, the only
free parameters left in the NSA model are the distance and temperature. TA05 provide a constraint on the distance to \tb\, depending on the mass
of the secondary.  We investigated fits for a range of masses/distances from 0.1 \msun (1.0 kpc) to 1.2\msun (2.3 kpc).  Statistically
acceptable fits are found for all distances, although in all cases only for a higher column $\simgt10^{20}$ cm$^{-2}$.  However, especially for
$d=1$ kpc, but also for the longer distances, the effective temperature of the NSA is low, and in all cases it contributes merely \til10\% of the flux, even
including energies down to 0.01 keV.  If one performs an $F$-test to compare the best fit PL (which has a low column) to the best fit PL+NSA
(which requires a higher column), the low value indicates that there is no need for the inclusion of the soft component to model the data; but of
course this all depends on the actual column.

In general, the quality of the fits is fairly similar and each yields an unabsorbed 0.01--10 keV X-ray luminosity of $5\times10^{-13}$\ergsqcmsec.  Although two-component models yield slightly better fits, in most cases consideration of the
parameter values indicate problems: extreme plasma temperatures or unusually low/high temperatures for the neutron star (either NSA or
blackbody). We defer further discussion to \S~\ref{sect:1023disc}.

\section{\ta}
\begin{deluxetable*}{llllllll}
%\tablenum{1}
\tablewidth{0pt} 
\tablecaption{Optical eclipse timings and $O-C$ results for \ta. \label{tab:0932eph} } 
\tablehead{\colhead{HJD(TT)} & \colhead{$\sigma({\rm HJD})$} & \multicolumn{2}{c}{$P=0.0663035(4)$ d}&\multicolumn{2}{c}{$P=0.0661618(4)$d} &\multicolumn{2}{c}{$P=0.0660206(5)$d} \\
 \colhead{2453000 +}& \colhead{($\times10^{-4}$)}&\colhead{Cycle} & \colhead{$O-C$}& \colhead{Cycle} & \colhead{$O-C$}& \colhead{Cycle} & \colhead{$O-C$}\\
&& \colhead{ Count}&\colhead{(cycles$\times10^{-5}$)}&\colhead{ Count}&\colhead{(cycles$\times10^{-5}$)}&\colhead{ Count}&\colhead{(cycles$\times10^{-5}$)}
}
\startdata
106.68445 & 4.5 &   1 &$3.3\pm6.8$  &   1  &$0.0\pm6.8$    &   1& $-3.2\pm6.8$        \\
106.75044 & 2.0 &   2 &$-1.4\pm3.0$  &   2  &$-2.5\pm3.0$   &   2& $-3.7\pm3.0$        \\
106.81690 & 2.8 &   3 &$1.0\pm4.0$  &   3  &$2.0\pm4.2$    &   3& $3.0\pm4.2$         \\
137.64816 & 1.6 & 468 &$2.9\pm2.4$  & 469  &$0.5\pm2.4$    & 470& $-1.9\pm2.4$        \\
137.71423 & 1.4 & 469 &$-0.5\pm2.1$  & 470  &$0.8\pm2.1$    & 471& $-1.0\pm2.1$        \\
137.78013 & 2.3 & 470 &$-6.6\pm3.5$  & 471  &$-4.7\pm3.5$   & 472& $-2.8\pm3.5$        \\
137.8470  &   3 & 471 &$1.4\pm4.5$   & 472  &$5.5\pm4.5$    & 473& $9.6\pm4.5$        \\
\enddata
\end{deluxetable*}
\subsection{Light Curves}
\begin{figure*}[!tb]
%\resizebox{.95\textwidth}{!}{\rotatebox{0}{\plotone{0932_oplcs.eps}}}
\resizebox{.95\textwidth}{!}{\rotatebox{0}{\plotone{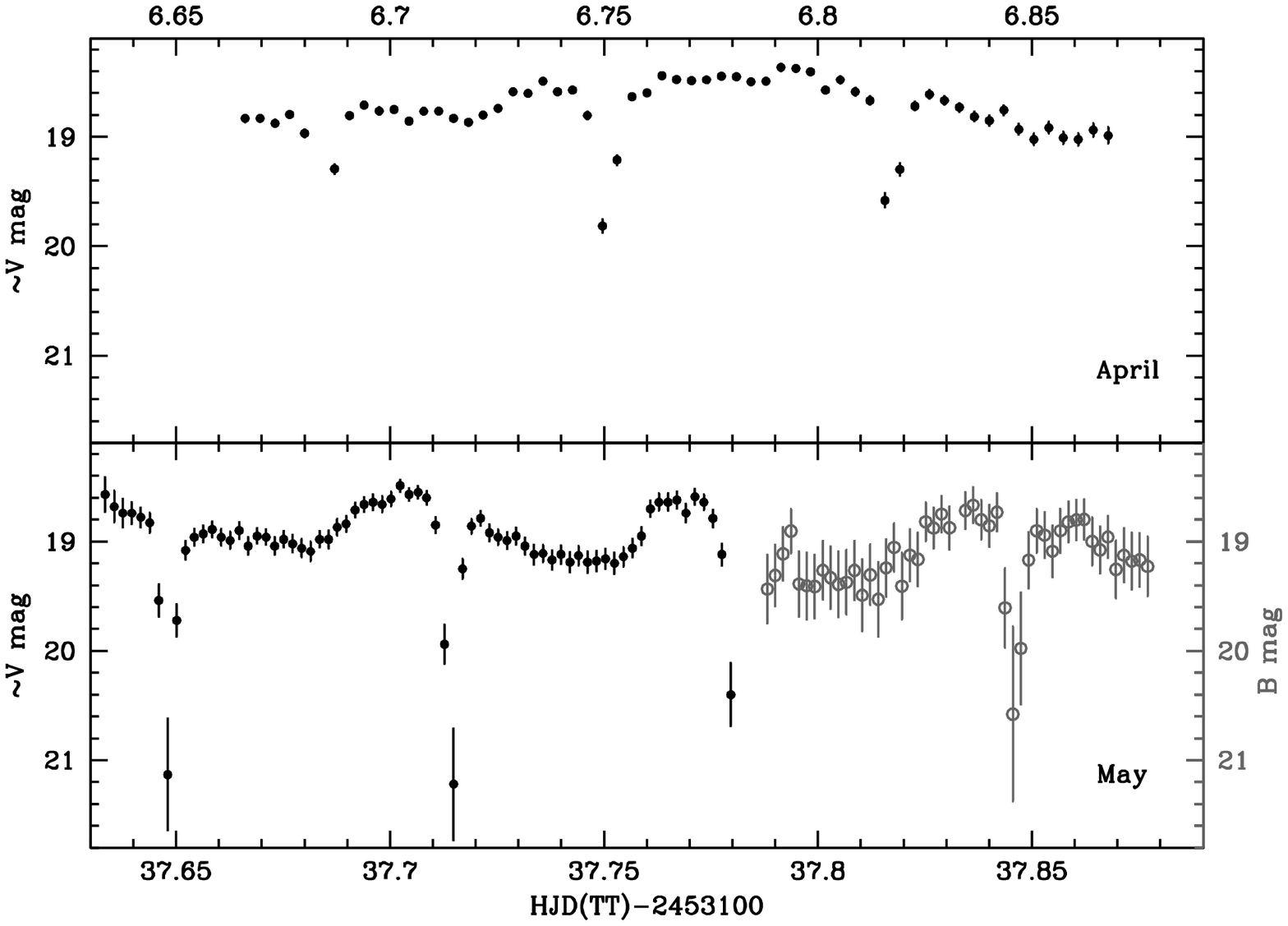}}}

\caption{Optical light curves for \ta. Black filled circles (and error bars) show data from the NOFS 1-m telescope.  These data were taken in white
  light; the $\sim V$ indicates that an approximate zero-pointing onto the Johnson system has been applied. Dark grey open circles show the $B$-band
  data from \sat/OM. Seven eclipses are clearly visible, as well as, non-eclipse modulation structure that changes significantly between the
  epochs. \label{fig:0932lcs}}
\end{figure*}

Owing to the low X-ray count rate of this source it was not possible to construct light curves with any useful time resolution.  However, at $B\sim19$
the OM light curve could usefully be binned to 160s, similar to the time-resolution of the NOFS data.  We present the three light curves from 2004
April and May in figure~\ref{fig:0932lcs}. \begin{figure}[!tb]
%\resizebox{.47\textwidth}{!}{\rotatebox{0}{\plotone{0932_OCplots.eps}}}
\resizebox{.47\textwidth}{!}{\rotatebox{0}{\plotone{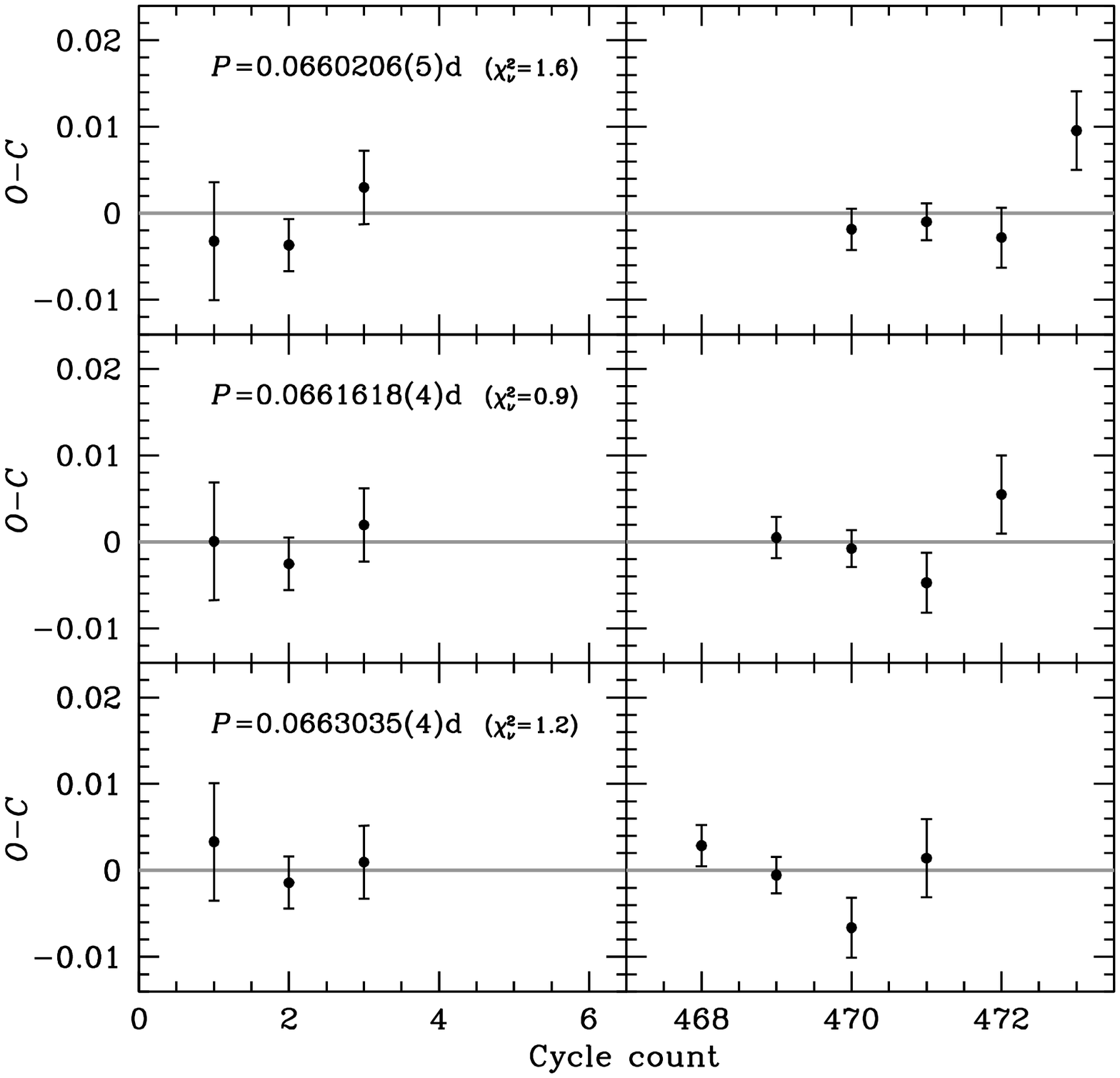}}}

\caption{$O-C$ plots for the eclipse timings of \ta. The best fit is for $P= 0.0661618$d but it is clear that a cycle count ambiguity of one between
  April and May observations makes little difference to the fits. \label{fig:0932oc}}
\end{figure}
 Consistent with the double-peaked optical spectra, this source is clearly at a high-inclination to our
line of sight, exhibiting deep, 2 magnitude eclipses.  We have undertaken an $O-C$ analysis using the 7 eclipse times spanning roughly a
month. From the 4 May points alone we can determine the period to within 0.2\%.  Unfortunately, this is not quite high enough precision to
facilitate an unambiguous cycle count back to the April epoch.  The best fit solution yields:
\[ {\rm HJD(TT)_{min}}= 2453122.2324(1) + E* 0.0661618(4) {\rm d}.\] 

In table~\ref{tab:0932eph} and fig.~\ref{fig:0932oc} we present our full \linebreak $O-C$
results for the 3 candidate solutions.  Clearly, additional photometry will be needed to confirm the correct one.

\begin{figure*}[!tb]
%\resizebox{.95\textwidth}{!}{\rotatebox{0}{\plotone{0932_opfds.eps}}}
\resizebox{.95\textwidth}{!}{\rotatebox{0}{\plotone{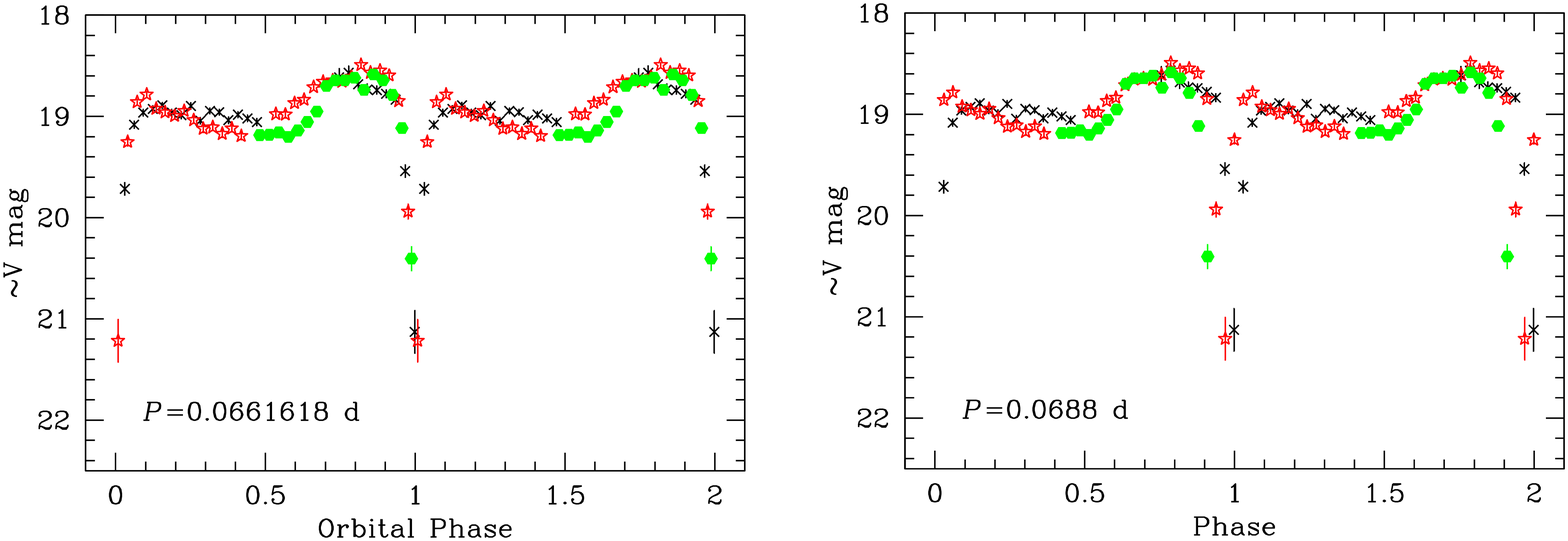}}}

\caption{Optical light curves for \ta\ from the 2005 May 13 observations.  The data have been folded on: ({\it left panel}) the orbital period, as determined from
  eclipse timings; ({\it right panel}) a longer period of 0.0688 d, as given by a periodogram analysis of the out-of-eclipse modulation. Each
  cycle has been marked distinctly by different symbols (and colors- electronic edition only) to aid the eye.\label{fig:0932fds}}
\end{figure*}

\ta\ exhibits variability in addition to its eclipses, which is notably different between the two epochs.  In 2005 April there appears a long
term trend, a slow rise and decline, with no obviously repeatable modulation on the orbital period.  In contrast, in 2005 May, there is no long
timescale change in flux, but there is a clear periodic smooth variation.  Close inspection of the light curves in the lower panel of
fig.~\ref{fig:0932lcs} reveals a possible trend in the appearance of the pre-eclipse hump: it appears slightly later in phase each cycle.  We fit a
sinusoid plus first harmonic model to all the non-eclipse datapoints, this gives a best fit period of $0.0688^{+0.0012}_{-0.0006}$ d, which is
3.8\% longer than the orbital period, hence 
nominally discrepant at the 4$\sigma$ level. Again with merely four cycles of data, this period is far from conclusive.  In
fig.~\ref{fig:0932fds} we show the NOFS data folded on both the orbital period and this longer one. The obvious explanation for such a period a
few percent longer than the orbital is the superhump effect \citep[see][for a detailed discussion of the superhump phenomenon]{patt98}. 
\begin{deluxetable*}{lllllll}
%\tablenum{1}
\tablewidth{0pt} \tablecaption{X-ray Spectral Fits for \ta\label{tab:0932xfits} } 
\tablehead{ \colhead{Model} &\colhead{Goodness}\tablenotemark{a} & \colhead{$N_H$} & \multicolumn{2}{c}{Partial covering}& \colhead{kT (keV)}&\colhead{Flux\tablenotemark{b} }\\
 & \colhead{of Fit} & \colhead{(cm$^{-2}$)}&$N_H$ (cm$^{-2}$) & frac. & \colhead{or $\Gamma$}& }
\startdata

  bremss & 70\% & $6^{+2}_{-1}\times10^{22}$&\nodata&\nodata& 200 (pegged) &$2.4\pm0.8$\\
  & 56\% & $1.39\tablenotemark{c}\times10^{20}$ & $10^{+5}_{-3}\times10^{22}$ & $0.95\pm0.02$& 200 (pegged) &$1.5\pm0.7$\\
  & 61\% & $1.39\times10^{20}$ & $10^{+5}_{-3}\times10^{22}$ & $0.95\pm0.02$& 30  &$1.5^{+1.6}_{-0.3}$\\
  PL & 76\% & $3\pm3\times10^{22}$ &\nodata&\nodata& $0.5\pm0.7$ &---\tablenotemark{d}\\ 
  & 59\% & $1.39\times10^{20}$ & $9\pm5\times10^{22}$ & $0.9\pm0.1$ & $0.8\pm0.6$&$1.7\pm0.6$\\

\enddata
\tablenotetext{a}{As the fitting utilized Cash statistics, a Monte Carlo method was used to find the percentage of
  simulated spectra based on the parameter space of the model fit had $C$-statistic values less than than of the fit to the data.  A good fit
  should have a value around 50\%.}
\tablenotetext{b}{Unabsorbed flux in the 0.01--10 keV range, in units of $10^{-13}$\ergsqcmsec}
\tablenotetext{c}{Model parameters without quoted uncertainties have been fixed at the value given.}
\tablenotetext{d}{The singly absorbed power law is such an unphysical fit that flux estimates were meaningless.}
\end{deluxetable*}

\subsection{X-ray Spectra}
Even after our application of {\tt epreject}, the background is estimated to contribute 45\% of the pn flux within our source aperture (only
$\simlt15$\% for the MOS detectors).  With 50 and 55
counts total in the pn and MOS detectors respectively, the use of smaller 5cts/bin binning and Cash statistics is the best approach.  First,
we examined and fit the background spectra.  For this we could bin to 20cts/bin and still have good spectral resolution, and we simply aimed to find
the best empirical model to represent these contributions.  Our final model consists of a twice-broken power law (i.e. 3 different indices), plus
Gaussian lines to account for the various metal emission lines.  We performed a joint fit to MOS background spectra, and independently for the
pn.  Next, we used this model (with relative normalisations scaled according to extraction area) as the fixed background component in our
source+background fitting.

\begin{figure}[!tb]
%\resizebox{.47\textwidth}{!}{\rotatebox{0}{\plotone{0932_xspec_fit.eps}}}
\resizebox{.47\textwidth}{!}{\rotatebox{0}{\plotone{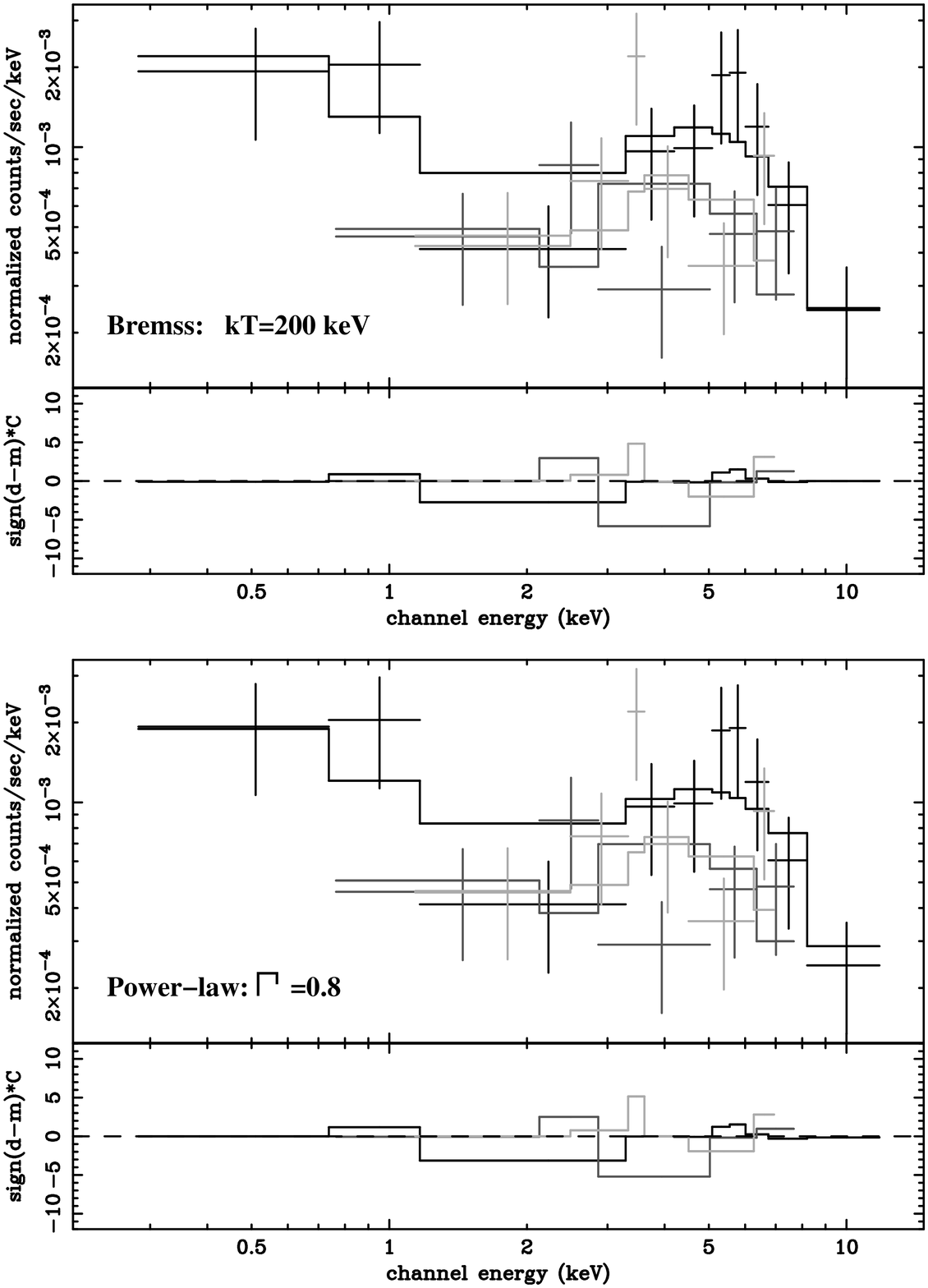}}}
\caption{\sat/EPIC spectra for \ta: {\it black}--pn; {\it dark grey}--MOS1; {\it light grey}--MOS2.  {\it Top}: best fit ``CV'' model
  consisting of a single temperature bremsstrahlung, with partial covering absorption (see text for details). {\it Bottom}: best
  fit ``quiescent LMXB'' -- power law emission model instead. 
\label{fig:0932xspec}}
\end{figure}

With our poor statistics we concentrated on finding the simplest model fit to the data, starting with a bremsstrahlung plus a single absorption
(i.e. for Galactic) column.  First, we found that there is significant column in excess of that estimated for the line-of-sight, our best fit
requiring $N_H=5.6^{+2.0}_{-1.3}\times10^{22}$ cm$^{-2}$, as compared to the dust-map estimate (from the {\tt nH} tool) of $N_H=1.39\times10^{20}$ cm$^{-2}$.  Moreover, the
temperature pegged at the 200 keV limit, well in excess of a physically plausible fit for a CV, indicating that this model is simply not a
good representation and perhaps a power law would be equally good.  The ``best'' (physically plausible) fit in this case gave
$\Gamma=0.5\pm0.7$ with $N_H=3\pm3\times10^{22}$ cm$^{-2}$; generally poorly constrained and a poor goodness of fit of 76\%\footnote{This XSPEC
  command simulates a user defined number of spectra (we used 5000) based on the model and writes out the percentage of these simulations with
  the fit statistic less than that for the data. If the observed spectrum was produced by the model then this number should be around
  50\%. This command only works if the sole source of variance in the data is counting statistics.}.  For high-inclination LMXBs, it is often
found that there is additional local absorption due to material vertically extended at the accretion disc rim \citep{chur01}, while for IPs the
material in the accretion flow can have the same effect.  In both cases, the absorption may only partially obscure the X-ray source spatially
and, for the case of an orbitally averaged spectrum as we have here, only for part of an orbit.  Hence, we included an additional column scaled by its
partial covering  fraction, whilst fixing the other at the nominal (maximum) Galactic value.  The inclusion of this extra component does improve
both fits, increasing the photon index, but the plasma temperature still pegs at its maximum.  At the same time, if we fix $kT_{\rm br}=30$keV
(typical for IPs) this
only increases goodness from 56\% to 61\%. The covering fraction is in fact close to unity ($0.9\pm0.1$ and $0.95\pm0.02$ respectively).  In
conclusion, our joint fits to the X-ray spectra (as shown in fig.~\ref{fig:0932xspec}, and detailed in Table.~\ref{tab:0932xfits}) can only constrain the absorption column, requiring a significant
local absorption, which could equally be accounted for in a mCV interpretation or (high inclination) LMXB interpretation.  As to the emission
model, the data seem best fit with a power law, but this is far from conclusive.

\section{Nature of \tb}
\label{sect:1023disc}
%In this section, we will review how the observational evidence now stands versus the two favoured models:  LMXB or magnetic CV.
%\subsection{LMXB?}
The observational constraints supplied by our new optical and especially X-ray data lend significant support to the LMXB hypothesis of
TA05. The 0.5-10 keV luminosity of \tb\ is $10^{32.4}$\ergsec\ assuming a distance of 2 kpc, which places it squarely in the range of luminosities
for neutron star LMXBs in quiescence \citep[see e.g.][]{camp04b}.  However, in this energy range the flux is dominated (97\%) by the hard power law
component. Recently, \citet{jonk04a,jonk04b} examined the relation between the X-ray luminosity of neutron star transients in quiescence and the
fractional contribution of the power law component.  They found a correlation for $L_X\simgt1-2\times10^{33}$\ergsec, and an anti-correlation
for lower luminosities. With its relatively low luminosity \tb\ lies at the top end of the anti-correlation having a negligible NSA
contribution, similar to both the LMXB in Terzan 5 and the millisecond pulsar SAX J1808.4-3658 in quiescence.  In the pulsar, the leading model for the
power law component is the interaction of the pulsar wind with that from the donor star \citep{stel94,burd03}, but as yet there is no clear
origin for the case of
weakly magnetized neutron stars.  Since \tb\ has not been observed with a large area X-ray detector, such as the {\em RXTE}, during an outburst,
we have no constraints on its
pulsed X-ray emission; it is interesting to speculate that it too may host a  millisecond pulsar.

 Another notable feature of \tb\ is its short timescale (i.e. \til10's of minutes) X-ray variability; again this is somewhat unusual, but not without
precedent, for example Cen X-4 and EXO 1745-248  \citep{camp04a,wijn05}.  As for the spectral parameters themselves: with $kT_{\rm eff}\approx50$ eV it is lower than for most
quiescent LMXBs (i.e. 100--300 eV), but fits in with the trend for those systems with very weak NSA components; the photon index$\approx1.3$ lies well
within the normal 1--2 range. 

 Lastly, now that we have a measure of the quiescent X-ray luminosity, we can consider how much brighter it may have
become during its active state (outburst), and consider its detectability during previous X-ray observations. A search of the HEASARC archives
yields no previous pointed observations by other X-ray satellites. Naturally, its position was covered by both the {\em ROSAT} All-Sky
Survey (RASS) and the {\em RXTE}/ASM, though in neither has it been detected. The null result for RASS (upper limit of 0.03 counts/s) is consistent with our \sat\ flux measures: we
predict merely 0.015 counts/s in the RASS 0.5--2.5 keV band.  However, the fact that \tb\ remained undetected by the ASM even during its bright
phase can provide a useful constraint.  The nominal  ASM point source sensitivity is  $8.4\times10^{-10}$\ergsqcmsec, hence \tb\ could have brightened by a factor \til2000
and still remained undetected by the ASM.  Such a brightening is well-within the wide spread seen amongst X-ray transients.  

In addition to the X-ray, we can examine its activity history via a number of optical observations.  Catalog data provide snapshots of its state
at their epochs of observation.  We have four independent data points spanning 1952 to 2000, and in each case the approximate $B$ and $R$
magnitudes show the source to be quiescent.  Furthermore, all observations since the late 2000/early 2001 epoch of the Sloan spectroscopy and
the Bond et al. series of observations show the source to be quiescent.  It therefore appears that \tb\ has a low-duty cycle of activity, which is
seen in  many other X-ray transients, and moreover, fits in with a low-temperature and luminosity from the neutron star atmosphere. 

In summary, we can argue that all observations to date find a system with characteristics and behaviour fully consistent with a LMXB, which
spends most of its time in quiescence. 

%\subsection{mCV?}
For the CV alternative, we immediately run into problems: what provides its X-ray emission in such a quiescent state?  A number of CVs are observed to enter very low
states, optically up to \til7 mags fainter than normal: the VY Scl stars.  However, they spend the majority of their time in active states. Only one IP has been observed to transition into such a
very low-state, V1223 Sgr \citep{garn88}, in a survey of the Harvard plate collection, hence nothing more is known about its
characteristics in that state.  Physically, in the absence of mass transfer
onto a white dwarf, the only plausible source of X-rays appears to be the chromospheric emission of the fast-rotating donor: as is the presumed case for
SDSS~J155331.12+551614.5 \citep{szko04a}.  But for \tb, this option is inconsistent with the hardness of its X-ray spectrum.  All active binaries when fit
with optically thin thermal emission models yield temperature $\simlt2$ keV \citep[see results from Chandra observations of 47 Tuc and from the
  RASS:][]{hein05,demp93}, not $\simgt15$ keV as we find.  Moreover, the ratio
of X-ray to optical flux is far greater than for any of the RASS or 47 Tuc active binaries.

\section{Nature of \ta}
Although  the rough $V$ magnitudes of
\ta\ (6 epochs- SDSS photometry through \sat\ observations) do vary, they differ by  much less than a magnitude, which is likely due to random
sampling of the eclipses in some of the longer exposures. With no clear evidence of significant state changes (unlike \tb) a transiently accreting LMXB
(or dwarf nova) model seems less favored.  In any case the high excitation lines of its optical spectra suggest either an active LMXB or magnetic CV. Furthermore,  the high X-ray
column (and partial covering model) further restrict the options to either a high-inclination LMXB or most likely an IP.  As we noted, from the
X-ray spectral fits alone it is hard to choose between these competing interpretations.  Hence, we have also compared the  X-ray to optical flux ratio of  \ta\ to those of all known IPs
\footnote{See Koji Mukai's compendium, http://lheawww.gsfc .nasa.gov/users/mukai/iphome/iphome.html} and to a sample of 8 high inclination LMXBs
with periods less than 10 hrs. Clearly, the exact details of X-ray spectrum, and the Galactic absorption are not taken out in these
estimates.  Nevertheless, if we normalize our values such that \ta\ has $F_X/F_{opt}=1$ we find it fits well within the scatter of values for IPs, which
range from 0.1 to 10, though clustered in the 1--2 range.  In contrast, most of the  high inclination (actively accreting) LMXBs have $F_X/F_{opt}\sim100-1000$; low for LMXBs because the disk obscures much of the
direct X-ray flux, but still orders of magnitude higher. 

The absence of a firm detection of additional periodicities in the optical light curve leaves the IP candidacy of \ta\ very much a matter
of debate.  There is also our tentative detection of a photometric variation a few percent longer than the orbital period, which would be
difficult to explain in an IP.  Much more extensive optical photometric monitoring is without doubt required to resolve these two issues.

\section{Conclusions}
We have obtained \sat\ X-ray and additional optical followup observations of two accreting binaries recently discovered in the SDSS: \tb\ and
\ta. Consideration of their optical spectra and magnitudes (and history thereof) suggests that we observed \tb\ during its (normal) inactive state, while
\ta\ was actively accreting at the time. 

For \tb\ the X-ray results favor the (transient) LMXB hypothesis of \citet{thor05}, both in terms of X-ray luminosity, spectral parameters and the
relative contributions of the best fit neutron star atmosphere + power law model.  The distance independent X-ray-to-optical flux ratio and
the estimated range of X-ray luminosities are inconsistent with what we would expect from an inactive CV, i.e. the chromospheric emission from
the low-mass donor star. 

Our two epochs of
time-series optical photometry confirm that \ta\ is indeed a high inclination system, exhibiting deep eclipses on a period of 1.57 hr. Unfortunately, given the very
low X-ray count rate of \ta\ we are unable to construct a useful X-ray light curve.  Neither are we able to constrain the emission model well
{\em but} our X-ray spectra clearly require significant local absorption.
This could be accounted for in either an Intermediate Polar or high inclination LMXB model.  We find that its X-ray-to-optical flux is fully
consistent with the known IPs, but at least an order of magnitude less than that of the  X-ray faintest LMXB.

\acknowledgments
We thank the anonymous referee for their timely and helpful report. This work was supported by \sat\ grant NNGO4GG66G to the University of
Washington and is based on observations obtained with \sat, an ESA science mission with instruments and contributions directly funded by ESA
Member States and the USA (NASA).  G. Schmidt acknowledges the support of NSF grant AST 03-06080.

    Funding for the creation and distribution of the SDSS Archive has
been provided by the Alfred P. Sloan Foundation, the Participating
Institutions, the National Aeronautics and Space Administration, the
National Science Foundation, the U.S. Department of Energy, the
Japanese Monbukagakusho, and the Max Planck Society. The SDSS Web site
is http://www.sdss.org/.

    The SDSS is managed by the Astrophysical Research Consortium (ARC)
for the Participating Institutions. The Participating Institutions are
The University of Chicago, Fermilab, the Institute for Advanced Study,
the Japan Participation Group, The Johns Hopkins University, the Korean
Scientist Group, Los Alamos National Laboratory, the
Max-Planck-Institute for Astronomy (MPIA), the Max-Planck-Institute
for Astrophysics (MPA), New Mexico State University, University of
Pittsburgh, Princeton University, the United States Naval Observatory,
and the University of Washington.

%\thebibliography

%\bibliographystyle{/Users/homer/home/papers/bst/apj}
%\bibliography{/Users/homer/home/papers/bib/aas}

\begin{thebibliography}{25}
\expandafter\ifx\csname natexlab\endcsname\relax\def\natexlab#1{#1}\fi

\bibitem[{{Bond} {et~al.}(2002){Bond}, {White}, {Becker}, \&
  {O'Brien}}]{bond02}
{Bond}, H.~E., {White}, R.~L., {Becker}, R.~H., \& {O'Brien}, M.~S. 2002,
  \pasp, 114, 1359

\bibitem[{{Burderi} {et~al.}(2003){Burderi}, {Di Salvo}, {D'Antona}, {Robba},
  \& {Testa}}]{burd03}
{Burderi}, L., {Di Salvo}, T., {D'Antona}, F., {Robba}, N.~R., \& {Testa}, V.
  2003, \aap, 404, L43

\bibitem[{{Campana} {et~al.}(1998){Campana}, {Colpi}, {Mereghetti}, {Stella},
  \& {Tavani}}]{camp98}
{Campana}, S., {Colpi}, M., {Mereghetti}, S., {Stella}, L., \& {Tavani}, M.
  1998, \aapr, 8, 279

\bibitem[{{Campana} {et~al.}(2004){Campana}, {Israel}, {Stella}, {Gastaldello},
  \& {Mereghetti}}]{camp04a}
{Campana}, S., {Israel}, G.~L., {Stella}, L., {Gastaldello}, F., \&
  {Mereghetti}, S. 2004, \apj, 601, 474

\bibitem[{{Campana} \& {Stella}(2004)}]{camp04b}
{Campana}, S. \& {Stella}, L. 2004, Nuclear Physics B Proceedings Supplements,
  132, 427

\bibitem[{{Church}(2001)}]{chur01}
{Church}, M.~J. 2001, Adv. Sp. Res., 28, 323

\bibitem[{{Dempsey} {et~al.}(1993){Dempsey}, {Linsky}, {Schmitt}, \&
  {Fleming}}]{demp93}
{Dempsey}, R.~C., {Linsky}, J.~L., {Schmitt}, J.~H.~M.~M., \& {Fleming}, T.~A.
  1993, \apj, 413, 333

\bibitem[{{den Herder} {et~al.}(2001)}]{denH01}
{den Herder}, J.~W. {et~al.} 2001, \aap, 365, L7

\bibitem[{{Garnavich} \& {Szkody}(1988)}]{garn88}
{Garnavich}, P. \& {Szkody}, P. 1988, \pasp, 100, 1522

\bibitem[{{Heinke} {et~al.}(2005){Heinke}, {Grindlay}, {Edmonds}, {Cohn},
  {Lugger}, {Camilo}, {Bogdanov}, \& {Freire}}]{hein05}
{Heinke}, C.~O., {Grindlay}, J.~E., {Edmonds}, P.~D., {Cohn}, H.~N., {Lugger},
  P.~M., {Camilo}, F., {Bogdanov}, S., \& {Freire}, P.~C. 2005, \apj, 625, 796

\bibitem[{{Jonker} {et~al.}(2004{\natexlab{a}}){Jonker}, {Galloway},
  {McClintock}, {Buxton}, {Garcia}, \& {Murray}}]{jonk04b}
{Jonker}, P.~G., {Galloway}, D.~K., {McClintock}, J.~E., {Buxton}, M.,
  {Garcia}, M., \& {Murray}, S. 2004{\natexlab{a}}, \mnras, 354, 666

\bibitem[{{Jonker} {et~al.}(2004{\natexlab{b}}){Jonker}, {Wijnands}, \& {van
  der Klis}}]{jonk04a}
{Jonker}, P.~G., {Wijnands}, R., \& {van der Klis}, M. 2004{\natexlab{b}},
  \mnras, 349, 94

\bibitem[{Lewin {et~al.}(1995)Lewin, van Paradijs, \& Taam}]{lewi95}
Lewin, W. H.~G., van Paradijs, J., \& Taam, R.~E. 1995, in X-ray Binaries, ed.
  W.~H.~G. Lewin, J.~van Paradijs, \& E.~P.~J. van~den Heuvel (Cambridge:
  Cambridge University Press), p. 175

\bibitem[{{Patterson}(1998)}]{patt98}
{Patterson}, J. 1998, \pasp, 110, 1132

\bibitem[{Scargle(1982)}]{scar82}
Scargle, J.~D. 1982, \apj, 263, 835

\bibitem[{{Stella} {et~al.}(1994){Stella}, {Campana}, {Colpi}, {Mereghetti}, \&
  {Tavani}}]{stel94}
{Stella}, L., {Campana}, S., {Colpi}, M., {Mereghetti}, S., \& {Tavani}, M.
  1994, \apjl, 423, L47

\bibitem[{{Str{\" u}der} {et~al.}(2001)}]{stru01}
{Str{\" u}der}, L. {et~al.} 2001, \aap, 365, L18

\bibitem[{Szkody {et~al.}(2003)}]{szko03}
Szkody, P. {et~al.} 2003, \aj, 126, 1499

\bibitem[{Szkody {et~al.}(2004)}]{szko04a}
---. 2004, \aj, 128, 1882

\bibitem[{{Thorstensen} \& {Armstrong}(2005)}]{thor05}
{Thorstensen}, J.~R. \& {Armstrong}, E. 2005, \aj, 130, 759

\bibitem[{{Turner} {et~al.}(2001)}]{turn01}
{Turner}, M.~J.~L. {et~al.} 2001, \aap, 365, L27

\bibitem[{Warner(1995)}]{warn95}
Warner, B. 1995, Cataclysmic Variable Stars (Cambridge University Press), 57

\bibitem[{{Wijnands} {et~al.}(2005){Wijnands}, {Heinke}, {Pooley}, {Edmonds},
  {Lewin}, {Grindlay}, {Jonker}, \& {Miller}}]{wijn05}
{Wijnands}, R., {Heinke}, C.~O., {Pooley}, D., {Edmonds}, P.~D., {Lewin},
  W.~H.~G., {Grindlay}, J.~E., {Jonker}, P.~G., \& {Miller}, J.~M. 2005, \apj,
  618, 883

\bibitem[{{Woudt} {et~al.}(2004){Woudt}, {Warner}, \& {Pretorius}}]{woud04}
{Woudt}, P.~A., {Warner}, B., \& {Pretorius}, M.~L. 2004, \mnras, 351, 1015

\bibitem[{{York} {et~al.}(2000)}]{york00}
{York}, D.~G. {et~al.} 2000, \aj, 120, 1579

\end{thebibliography}

\end{document}